\documentclass{aa}
\usepackage[applemac]{inputenc}
\usepackage{natbib}
\usepackage{txfonts}
\usepackage{graphicx}
\usepackage{color}
\bibpunct{(}{)}{;}{a}{}{,}

\begin{document}

\title{On the stability of non--isothermal Bonnor--Ebert spheres}
\author{O. Sipilä
\and{J. Harju}
\and{M. Juvela}
}
\institute{Department of Physics, PO Box 64, 00014 University of Helsinki, Finland\\
e-mail: \texttt{olli.sipila@helsinki.fi}
}

\date{Received / Accepted}

\abstract
{}
{We aim to derive a stability condition for non-isothermal Bonnor--Ebert spheres and compare the physical properties of critical non-isothermal and isothermal gas spheres. These configurations can serve as models for prestellar cores before gravitational collapse.}
{A stability condition for non--isothermal spheres is derived by constructing an expression for the derivative of boundary pressure with respect to core volume. The temperature distribution is determined by means of radiative transfer calculations. Based on the stability analysis, we derive the physical parameters of critical cores for the mass range $0.1 - 5.0\,M_{\odot}$. In addition, the properties
of roughly Jupiter-mass cores are briefly examined.}
{At the low-mass end the critical non-isothermal sphere has lower central density and a slightly larger physical radius than the corresponding isothermal sphere (i.e. one with the same mass and average temperature). The temperature decrease towards the core centre becomes steeper towards smaller masses as the central density becomes higher. The slope depends on the adopted dust model. We find that the critical dimensionless radius increases above the isothermal value $\xi_0=6.45$ for very low-mass cores ($< 0.2 M_\odot$). However, in the mass-range studied here the changes are within 5\% from the isothermal value.}
{The density structures of non-isothermal and isothermal Bonnor--Ebert spheres for a given mass are fairly similar. However, the present models predict clear differences in the average temperatures for the same physical radius. Especially for low-mass cores, the temperature gradient probably has implications on the chemistry and the observed line emission. We also find that hydrostatic Jupiter-mass cores with radii less than 100\,AU would have very high boundary pressures compared with typical pressures in the interstellar space.}


\maketitle

\section{Introduction}

Physical conditions prevailing in prestellar cores, i.e. gravitationally bound concentrations of molecular gas, constitute the initial set-up of star formation. Prestellar cores are generally not spherical in shape, but it has been found that a Bonnor--Ebert (BE) sphere \citep{Bonnor56, Ebert55}, i.e. an isothermal gas sphere in hydrostatic equilibrium, can succesfully approximate several prestellar cores \citep[e.g.][]{Bacmann00, Alves01, Kandori05}. While the density structure predicted by a BE model can give a good fit to observations, there may be problems with e.g. the pressure required for the core to be stable \citep[see e.g. the review by][and the references therein]{Bergin07}. Furthermore, the isothermal nature of the observed cores cannot be ascertained, and it is possible that at least some of the ``isothermal'' cores possess a significant temperature gradient due to attenuation of starlight \citep[e.g.][]{Zucconi01, Ward-Thompson02, Pagani04, Crapsi07}. A hydrostatic core with the temperature decreasing inwards is one possible configuration preceding the gravitational collapse, and can thus serve as the starting point for dynamical models.

In light of the above, it is of interest to study what kind of constraints could be imposed on the stability of a non--isothermal version of the Bonnor--Ebert sphere (hereafter BES). The non--isothermal (``modified'') Bonnor--Ebert sphere (hereafter MBES) has been discussed previously by e.g. \citet{Evans01}, \citet{Galli02}, \citet{Keto05} and \citet{Sipila10}. In this paper, we derive an explicit stability condition for the MBES, using a method similar to that of \citet{Bonnor56} in his derivation of the isothermal stability condition. The non--isothermal stability condition will allow us to calculate the critial radius for the MBES, and this is turn yields information on e.g. the possible values of the density contrast and boundary pressure for critical modified BE spheres.

The paper is structured as follows. In Sect.\,\ref{Sect2}, we discuss the properties of the MBES and present a method of obtaining the non--isothermal solution ($\xi$, $\psi$). We also present an explicit stability condition and discuss how the critical radius of an MBES of given mass can be found. In Sect.\,\ref{sect3}, we present the results of our analysis, i.e. the behavior of the physical characteristics (radius; density and temperature structure) of critical cores as a function of core mass. In Sect.\,\ref{sect4} we discuss the implications of the results presented in Sect.\,\ref{sect3}. In Sect.\,\ref{sect5}, we present our conclusions. Appendix\,\ref{appendixa} includes detailed derivations of some expressions presented in Section \ref{Sect2}, and Appendix\,\ref{appendixb} discusses thermodynamics of the MBES.

\section{The critical radii of non-isothermal Bonnor--Ebert spheres} \label{Sect2}

\subsection{The modified Bonnor--Ebert sphere}\label{sect2.1}

The Bonnor--Ebert sphere \citep{Bonnor56, Ebert55} is an isothermal hydrostatic gas sphere in hydrostatic equilibrium. By assuming the ideal gas equation of state and using the hydrostatic equilibrium condition, \citet{Bonnor56} obtained an expression for the density distribution of such a sphere \citep[eq. (2.3) in][]{Bonnor56}. The density distribution equation can be solved by expressing the density and radius in terms of non--dimensional variables $\xi$ and $\psi$ \citep[eq. (2.6) in][]{Bonnor56}; this leads to the Lane--Emden equation \citep[eq. (2.8) in][]{Bonnor56}, which reduces to a pair of first-order differential equations. The solution to the Lane--Emden equation then yields the physical parameters of the sphere by back-substitution.

By considering a radial dependence of the temperature, the above analysis can be generalized to non--isothermal spheres; this leads to a modified Bonnor--Ebert sphere \citep{Evans01, Keto05}. In this case, the density distribution equation takes the form
\begin{equation}\label{ddistr}
{1\over r^2} {d\over dr} \left( {r^2 \over \rho} \left[ T {d\rho \over dr} + \rho {dT \over dr} \right] \right) = - {4\pi Gm \rho \over k} \, .
\end{equation}
In \citet{Sipila10}, we discussed a modified version of Bonnor's substitutions \citep[eq. (2.6) in][]{Bonnor56} taking into account the radial dependence of the temperature:
\begin{equation}\label{eq2.1}
\rho = { \lambda \over \tau} {\mathrm e}^{-\psi}
\end{equation}
\begin{equation}\label{eq2.2}
R = \beta^{1/2}\lambda^{-1/2}\xi \, ,
\end{equation}
where $\tau = T/T_{\rm c}$, $T = T(r) = T(\xi)$, $T_{\rm c}$ is the temperature at the core centre, $\lambda$ is the mass density at the core centre and $\beta = kT_{\rm c}/4\pi Gm$, where $k$ is the Boltzmann constant and $m$ is the average molecular mass of the gas (here we assume that the gas consists of $\rm H_2$ and He, so that $m = 2.33$\,amu). We note that our substitutions are slightly different than those of \citet{Keto05}, who used the temperature at the core edge, $T_{\rm out}$, as the reference temperature. We have chosen the central temperature $T_{\rm c}$ as the reference temperature, because it depends less on the assumptions of the conditions outside the core. Now, for example, pressure and volume can be expressed in terms of $\xi$ and $\psi$ as
\begin{equation} \label{pressure}
p = {\rho kT \over m} = 4 \pi G \beta \lambda e^{-\psi},
\end{equation}
\begin{equation} \label{volume}
V = {4\pi\over3} r^3 = {4\pi\over3} \beta^{3/2} \lambda^{-3/2} \xi^3 \, ,
\end{equation}
where we used equations (\ref{eq2.1}) and (\ref{eq2.2}) and assumed the ideal gas equation of state $p = \rho kT/m$. We note that the dimension in the above equations is carried by $\beta$ and $\lambda$ (and of course by $G$), as $\xi$ and $\psi$ are dimensionless.

Inserting equations (\ref{eq2.1}) and (\ref{eq2.2}) into eq.\,(\ref{ddistr}), one obtains the non--isothermal (``modified'') Lane--Emden equation
\begin{equation}\label{eq2.3}
\xi^{-2} {d\over d\xi} \left[ \xi^2 \tau {d\psi \over d\xi} \right] = {1 \over \tau} {\rm e}^{-\psi} \, .
\end{equation}
Imposing the boundary conditions $\psi = 0$, $d\psi/d\xi = 0$, $\tau =1$ and $d\tau/d\xi = 0$ at the center one can (numerically) integrate eq.\,(\ref{eq2.3}) and thus obtain a density profile for the MBES, although the temperature profile must be calculated independently using, e.g., radiative transfer means. We discuss this point in further detail in the following Section.

Finally, we note that by choosing $T_c = T = $ constant ($\tau = 1$), one readily recovers the isothermal versions of equations (\ref{ddistr}) to (\ref{eq2.3}).

\subsection{Solving the non--isothermal Lane--Emden equation}\label{sect2.2}

Because of its temperature dependence, eq.\,(\ref{eq2.3}) cannot be readily solved. One thus needs to supply a temperature profile. In order to simulate the physical conditions inside prestellar cores, we have chosen to obtain the temperature profile using radiative transfer modelling of the dust component \citep{JP03, Juv05}. We assume in the following that $T_{\rm gas} = T_{\rm dust}$; this assumption is discussed below.

Before a temperature profile can be calculated, however, a first approximation density profile for the model core is needed. A natural starting point for the calculations is the BES, because the solution ($\xi$, $\psi$) to the Lane--Emden equation is known.

We first fix the mass of the core. The mass of an isothermal core can be written, using the variables introduced previously, as
\begin{equation}\label{eq2.6}
M_{\rm iso} = 4\pi \, \beta_{\rm iso}^{3/2} \lambda^{-1/2} \xi_{\rm out}^2 \, \biggl({d\psi \over d\xi}\biggr)_{\rm out} \, ,
\end{equation}
where $\beta_{\rm iso} = kT/4\pi Gm$. In the above and from this point on, the subscript ``out'' refers to a value taken at the edge of the core. Next, a value for the non--dimensional radius $\xi_{\rm out}$ has to be chosen. The choice of $\xi_{\rm out}$ determines the stability of the BES. The critical value is $\xi_0 \sim 6.45$; cores with $\xi_{\rm out} < \xi_0$ are stable, whereas cores with $\xi_{\rm out} > \xi_0$ are unstable \citep[see e.g.][]{Bonnor56}. Also, small values of $\xi_{\rm out}$ represent cores with a low central density and rather shallow density gradients, whereas cores with large $\xi_{\rm out}$ are centrally dense and present a large density gradient.

Having fixed $M$ and $\xi_{\rm out}$, we proceed to solve the central density $\lambda$ using eq.\,(\ref{eq2.6}). A density profile is then constructed using $\rho = \lambda \exp(-\psi)$, the isothermal counterpart of eq.\,(\ref{eq2.1}). A dust temperature profile for the core can now be calculated. In the temperature calculation, we assume that the core is embedded in a larger molecular cloud; we set $A_{\rm V} = 10$ at the edge of the core. The spectrum of the unattenuated interstellar radiation field is taken from \citet{Black94}. The final calculated temperature depends on the properties of the dust component. We carry out separate temperature calculations assuming two different types of grains, using grain opacity data from \citet[][henceforth OH94]{OH94} and from \citet[][henceforth LD01]{LD01}; the LD01 extinction curve was slightly modified as described in \citet{Sipila10}. In the latter case, we consider both silicate and graphite grains, with grain material densities 3.5\,g\,cm$^{-3}$ and 2.5\,g\,cm$^{-3}$, respectively. From now on, we will call core models with OH94 grains ``Type 1'' models, and core models with LD01 grains ``Type 2'' models. The OH94 and LD01 grain models were chosen for this study because these models describe grains in the dense medium inside prestellar cores, where grain properties and possibly their size distribution are thought to differ from those in the diffuse ISM \citep[e.g.][]{Steinacker10, Pagani10, Juvela2011b}.

In all calculations in this paper, we assume that $T_{\rm gas} = T_{\rm dust}$. This assumption should hold well for low mass cores and in the central parts of more massive cores ($M \sim 4-5\,M_{\odot}$; hereafter these will be referred to as ``high mass cores''), but generally in the outer parts of high mass cores $T_{\rm gas} ≠ T_{\rm dust}$ \citep[e.g.][]{Galli02, Keto05}. However, high mass cores should be approximately isothermal due to low average density \citep[see e.g.][]{Keto05}, and hence we can (qualitatively) expect the high mass MBES to behave like the BES.

After the temperature calculation, the modified Lane--Emden equation (eq.\,\ref{eq2.3}) is solved using the determined temperature profile. This yields the function $\psi$ as a function of $\xi$. The function $\psi$ is not yet self--consistent, however, since the solution was obtained starting from the isothermal core. We proceed by updating the central density of the core using the new function $\psi$. The mass of a non--isothermal core can be expressed as
\begin{equation}\label{eq2.7}
M_{\rm noniso} = 4\pi \, \beta_{\rm noniso}^{3/2} \lambda^{-1/2} \xi_{\rm out}^2 \, \tau_{\rm out} \, \biggl({d\psi \over d\xi}\biggr)_{\rm out} \, ,
\end{equation}
where now $\beta_{\rm noniso} = kT_c/4\pi Gm$. The mass of the core and the non--dimensional radius $\xi_{\rm out}$ are kept constant during the iteration, so that the above equation yields a new estimate of the central density $\lambda$. A new density profile is then calculated according to eq.\,(\ref{eq2.1}). A new temperature profile is resolved yielding a new solution to eq.\,(\ref{eq2.3}). The iteration is continued until the function $\psi$ converges; this happens typically after 3-4 iterations. When the iteration is complete, the physical parameters of the core can be derived using eqs.\,(\ref{eq2.1}) to (\ref{volume}).

\subsection{The critical radius of the MBES} \label{sect2.4}

As discussed in Sect.\,\ref{sect2.2}, the initial choice of $\xi_{\rm out}$ determines the density structure of the isothermal core, which one uses as a starting point for constructing the non--isothermal core. Because in the isothermal case the Lane--Emden equation does not depend on temperature, the solution function $\psi$ is always the same regardless of the choice of $\xi_{\rm out}$; the solution is, in this sense, universal. In the non--isothermal case the situation is different -- each value of $\xi_{\rm out}$ represents a unique core configuration with a unique temperature structure. This means in particular that the solution function $\psi$ depends on $\xi_{\rm out}$.

The stability condition of both the BES and the MBES can be found by constructing the derivative of the boundary pressure in terms of core volume. For a stable core, we expect pressure to increase in a contraction of the core, corresponding to a negative sign of the derivative. The critical point is the smallest value of $\xi_{\rm out}$ for which the pressure derivative turns positive ($\delta p / \delta V > 0$). As shown by \citet{Bonnor56}, {\sl all} values of $\xi_{\rm out}$ beyond this point represent unstable cores. We thus need to solve eq.\,(\ref{eq2.3}) for multiple values of $\xi_{\rm out}$, and look for the value of $\xi_{\rm out}$ for which the pressure derivative changes sign. In practice, when the iteration is complete, we extract the values of the parameters $\beta$, $\lambda$ and $\psi$\footnote{We extract the value of $\psi$ at the edge of the core, that is at $\xi_{\rm out}$, so that the mass of the core remains fixed.} corresponding to each $\xi_{\rm out}$ and combine them to form ``global'' functions $\beta(\xi_{\rm out})$, $\lambda(\xi_{\rm out})$ and $\psi(\xi_{\rm out})$. These will be needed to calculate the pressure derivative.

The pressure derivative can be constructed by using variational calculus; our method is analogous to that of \citet{Bonnor56}, with the exception that we have extended the discussion to non--isothermal models. In practice, the extension is carried out by considering variations of three parameters ($\beta$, $\lambda$ and $\xi$), instead of just the two latter ones needed in the isothermal analysis. In the case of the MBES, the pressure derivative takes the form (see also Appendix \ref{appendixa})
\begin{equation}\label{stability}
{\delta p \over \delta V} = {2 p \over 3V} { \left[ \beta^{-1} \delta\beta + \lambda^{-1} \delta\lambda - \left({d\psi \over d\xi}\right)_{\rm out} \delta\xi_{\rm out} \right] \over \left[ \beta^{-1} \delta\beta -  \lambda^{-1} \delta\lambda + 2 \, \xi_{\rm out}^{-1} \, \delta\xi_{\rm out} \right]  } \, .
\end{equation}
We insert into the above equation the global functions $\beta(\xi_{\rm out})$, $\lambda(\xi_{\rm out})$ and $\psi(\xi_{\rm out})$ and look for the value of $\xi_{\rm out}$ for which the pressure derivative changes sign; this value is the non--dimensional critical radius $\xi_1$. We have done this for a range of core masses; the results of the analysis are presented in Sect.\,\ref{sect3}.

The core mass and hence the total number of particles is conserved in a contraction of the core, i.e. $\delta N = 0$. This leads to a condition between $\delta\beta$, $\delta\lambda$ and $\delta\xi_{\rm out}$ (see Appendix \ref{appendixa}), which can be used to eliminate for example $\delta\beta$ in eq.\,(\ref{stability}). If this is done, it is straightforward to verify that in the isothermal limit, eq.\,(\ref{stability}) reduces to the isothermal pressure derivative \citep[eq. 2.16 in][]{Bonnor56}.

\section{Results}\label{sect3}

We have derived the physical parameters (radius, density, temperature) of critical MBESs for a range of core masses using the iterative approach described in Sect. \ref{Sect2}. In this Section we first compare the radial density distributions within a critical MBES and a critical BES, and thereafter describe how the "global" properties, i.e. the central density and the temperature profile of an MBES depend on the core mass.

\begin{figure}
\includegraphics[width=\columnwidth]{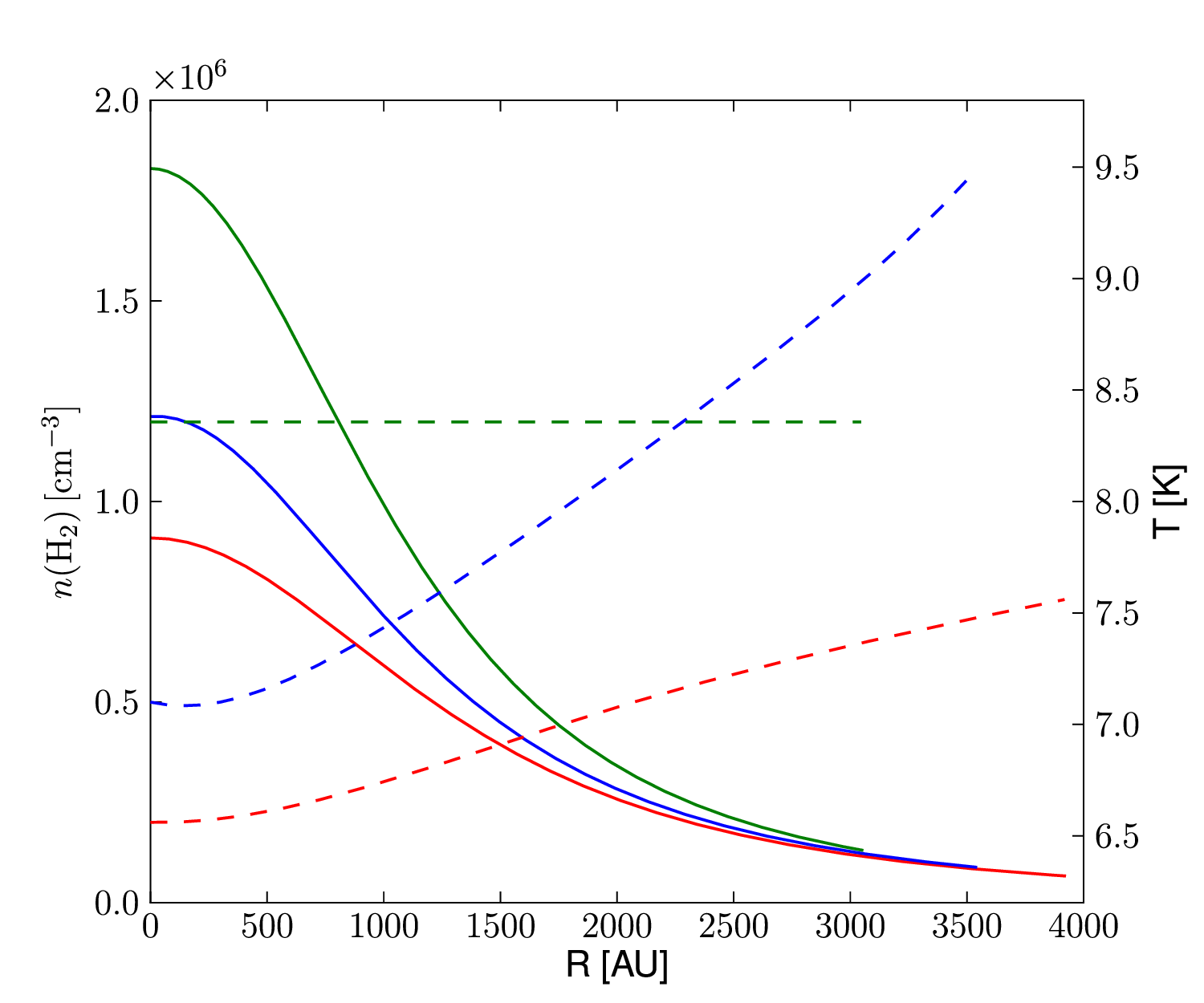}
\caption{The $\rm H_2$ number density (solid lines, left y--axis) of critical $0.25\,M_{\odot}$ Type 1 (OH94, blue) and Type 2 (LD01, red) MBESs as a function of radial distance from core centre. Also plotted is the density profile of the critical $0.25\,M_{\odot}$ BES corresponding to the mean temperature of the Type 1 MBES (green solid line; see text). The temperature profiles of the respective cores are plotted as dashed lines (right y--axis).}
\label{fig2}
\end{figure}

\subsection{Non--isothermal vs. isothermal: internal density structure} \label{sect3.4}

The density distributions of an MBES and a BES for a given mass are compared in Fig.\,\ref{fig2}. Here we plot the gas number density ($n({\rm H}_2)$, solid lines, left y-axis) as a function of radial distance from the core centre for a critical MBES of mass 0.25 $M_{\odot}$. The Type 1 core is plotted in blue and the type 2 in red. Also plotted (green solid line) is the density profile of a critical BES of the same mass and a temperature which is equal to the average temperature of the Type 1 core (cf. eq.\,\ref{tw}). The temperature profiles are plotted with dashed lines keeping the same color codes as for the densities (the scale is on the right).

On can see in the Figure that the higher central temperature of a BES as compared with an MBES permits also a higher central density, and the BES is more compact. A similar effect is seen between Type 1 and Type 2 cores.  The Type 1 core is slightly warmer and more centrally concentrated than the Type 2 core. The temperature difference between the centre and the outer boundary is clearly larger the Type 1 core (about 2.5\,K) than for Type 2 (about 1\,K). This is caused by the different dust models.

\subsection{The central densities and temperature profiles of critical non--isothermal cores} \label{sect3.3}

\begin{figure}
\includegraphics[width=\columnwidth]{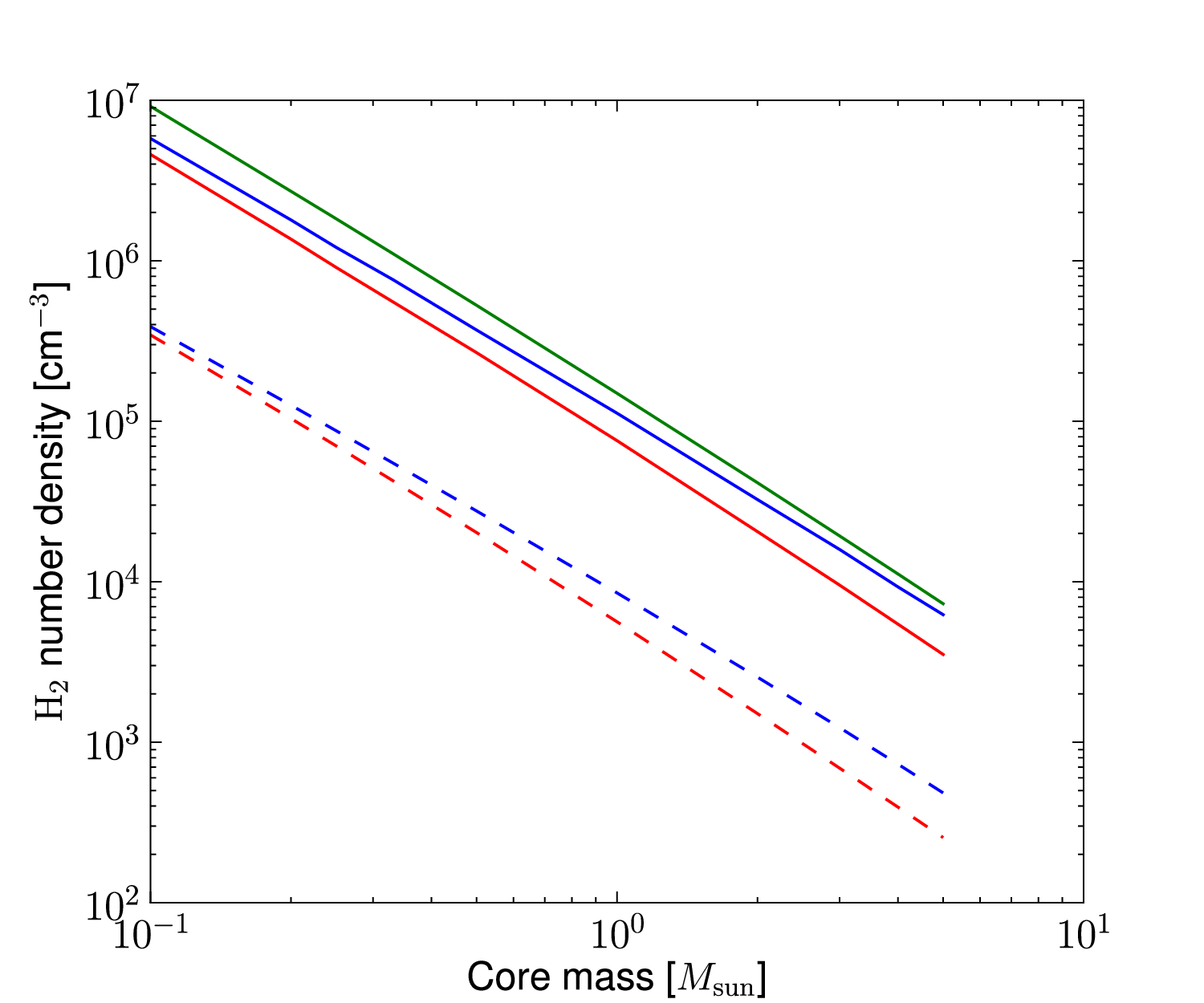}
\caption{The central H$_2$ number densities of Type 1 (OH94, blue solid line) and Type 2 (LD01, red solid line) critical MBESs as a function of core mass. The green solid line represents central densities of critical BESs corresponding to the mean temperatures of the Type 1 MBESs. Also plotted are the outer number densities of the respective cores types (dashed lines). }
\label{fig3}
\end{figure}

Figure \ref{fig3} plots the central number densities of critical MBESs and BESs as functions of the core mass. Also plotted are the MBES number densities at the outer boundary. The overall tendency seen in this Figure that the central density decreases steeply with the core mass can be understood from eqs. (\ref{eq2.6}) and (\ref{eq2.7}), according to which the central density (represented by the parameter $\lambda$) is approximately proportional to $M^{-2}$ when the non-dimensional radius $\xi_{\rm out}$ does not change.  The slight deviations from the $M^{-2}$ power-law are caused by temperature changes and small variations of $\xi_{\rm out}$ discussed below.  The central densities of BESs are higher than those of MBESs in the entire mass range, and the Type 1 cores are always denser than the Type 2 cores. The ratio of central and outer number densities of the MBESs remains approximately constant, regardless of core mass. There is however a small increase  in this ratio toward small core masses; we will discuss this in Sect. \ref{sect3.1}.

\begin{figure}
\includegraphics[width=\columnwidth]{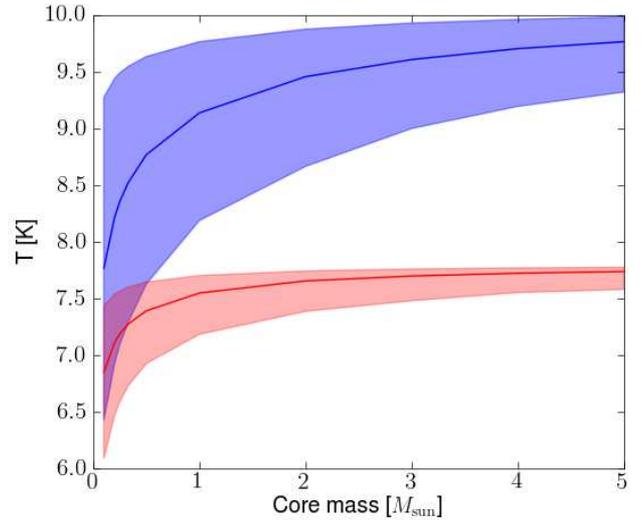}
\caption{The temperature structures of Type 1 (blue) and Type 2 (red) critical MBESs as a function of core mass. The upper bounds of the shaded areas represent the outer temperatures of the cores, the lower bounds represent central temperatures. The solid lines indicate mean temperatures (see text).}
\label{fig4}
\end{figure}

Figure \ref{fig4} plots the temperature structures of Type 1 (blue) and Type 2 (red) critical MBESs as a function of core mass. The upper bounds of the shaded areas represent the outer temperatures of the cores, the lower bounds represent central temperatures. Also plotted are the mean temperatures of the cores as defined by eq.\,(\ref{tw}) (solid lines). The temperature difference between the core centre and the edge increases towards smaller masses. This is caused by the increase in the central density (see Fig.\,\ref{fig3}) which leads to greater attenuation of external radiation. Low-mass MBESs have noticeable temperature gradients while cores with larger masses are almost isothermal (Type 1 cores still present a temperature gradient of $\sim$\,1\,K at $M = 5 \, M_{\odot}$). Looking at the mean temperatures, we see that for high mass cores the mean temperature approaches the outer temperature, indicating that most of the core mass lies in the outer parts of the core. For small core masses, the mean temperature approaches the central temperature, and the medium is more centrally concentrated.

\subsection{The critical non--dimensional radius $\xi_1$}\label{sect3.1}
\begin{figure}
\includegraphics[width=\columnwidth]{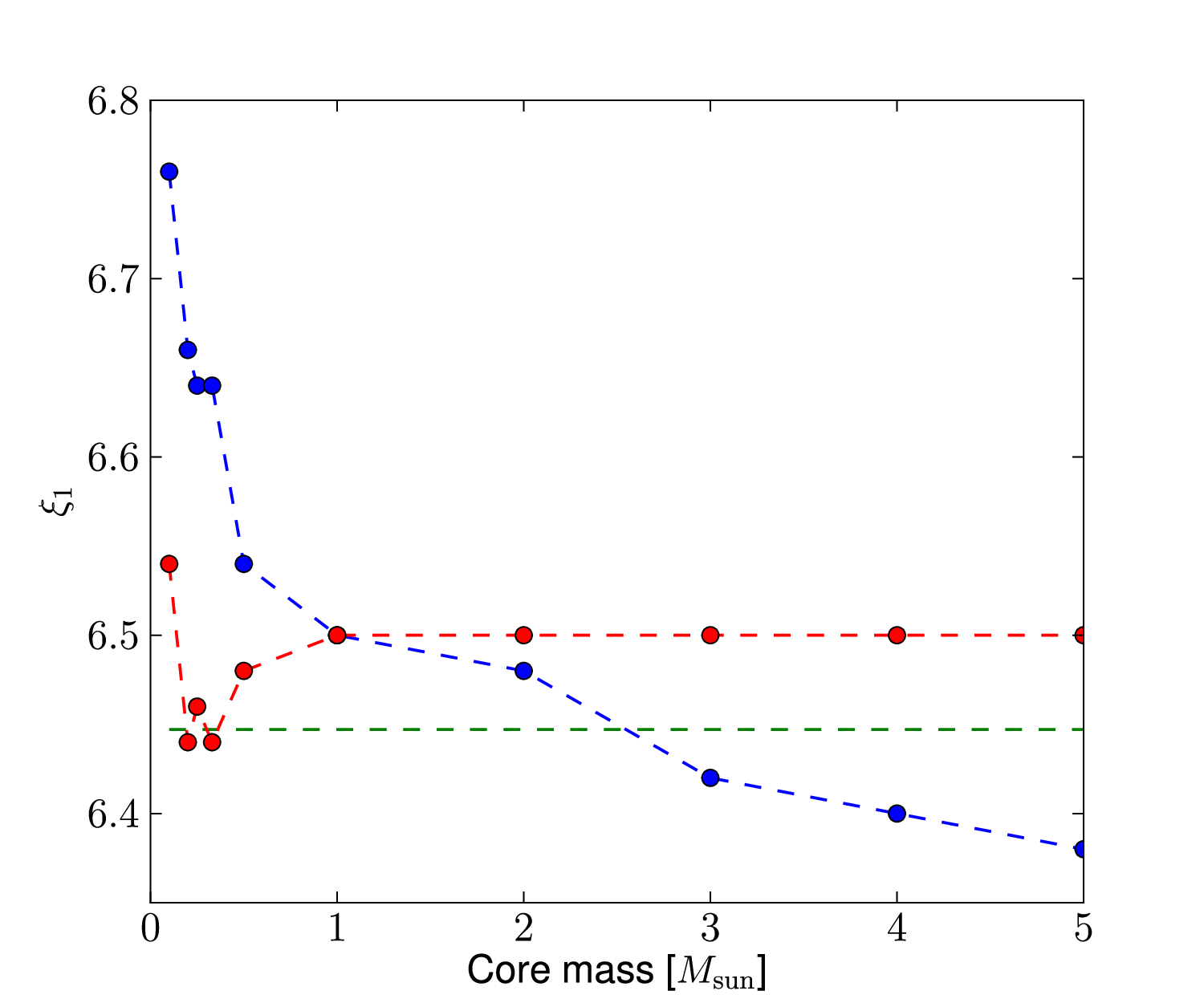}
\caption{The non--dimensional critical radius $\xi_1$ for Type 1 and Type 2 models (blue and red lines, respectively) as a function of core mass. Also plotted is the isothermal critical radius $\xi_0$ (green dashed line).}
\label{fig5}
\end{figure}

The non-dimensional outer radius $\xi_{\rm out}$ characterizes the shape of the density profile which can be compared with high-precision observations like in the famous case of B68 \citep{Alves01}. The critical value of $\xi_{\rm out}$ where an MBES becomes unstable against an increase in the external pressure is denoted here by $\xi_1$ (see Sect. \ref{sect2.4}). Figure~\ref{fig5} plots the value of $\xi_1$ calculated according to (eq.\,\ref{stability}) as a function of core mass. The blue and red lines correspond to Type 1 and Type 2 models, respectively. Also plotted in the Figure is the isothermal critical value $\xi_0 \sim 6.45$ (green dashed line), which does not depend on core mass.

One can notice here that for Type 1 cores $\xi_1$ increases clearly towards smaller masses, whereas for Type 2 cores $\xi_1$ is nearly constant and close to isothermal value $\xi_0$, except for the wiggle at the low-mass end. The different behavior of $\xi_1$ in the two cases is due to the different temperature structures of the two core types (Fig.\,\ref{fig4}). Consider first Type 1 cores. At high core masses, the temperature gradient of the critical core is $\sim 1$\,K, whereas for small core masses the gradient is $\sim 3.5$\,K. For Type 2 cores, the high mass temperature gradient is $\sim 0.3$\,K, whereas the low mass gradient is $\sim 1.0$\,K. Comparing with Fig.\,\ref{fig5}, we see that the critical radius correlates fairly well with the behavior of the temperature gradients; for Type 1 models, $\xi_1$ grows as the temperature gradient grows, and for Type 2 models, $\xi_1$ is approximately constant until the temperature gradient starts to grow toward small core masses. A more general conclusion is that small mass cores with high central density and steep temperature gradient are able to remain stable for slightly higher values of $\xi_1$.

The increasing temperature gradient towards lower masses is also reflected in the density contrast,
\begin{equation} \label{rho_grad}
\frac{\rho_{\rm c}}{\rho_{\rm out}} =
\frac{T_{\rm out}}{T_{\rm c}} \, e^{\psi_{\rm out}} \; .
\end{equation}
For MBESs the critical density contrast is generally close that of a critical BES ($\sim$ 14), but exceeds the isothermal value for the smallest masses.

We note that the values of $\xi_1$ plotted in Fig.\,\ref{fig5} are subject to numerical uncertainty, mostly due to Monte Carlo fluctuation in the temperature calculations, which can be up to 0.1\,K. This fluctuation is reflected on the pressure derivative (eq.\,\ref{stability}) through its dependence on $\beta$, $\lambda$ and $\psi$ and their numerical derivatives. To analyze the effect of numerical fluctuation, we performed the iterative calculations presented in Sect.\ref{Sect2} with multiple $\xi_{\rm out}$ grids of varying accuracy (meaning specifically the spacing between neighboring values of $\xi_{\rm out}$). It was found that the values of $\xi_1$ for a given core mass can change by up to 0.1 depending on the grid used, but the general shapes of the lines plotted in Fig.\,\ref{fig5} remain more or less the same in all cases studied. The numerical fluctuation is probably responsible for the small dip in the Type 2 data present in Fig.\,\ref{fig5}. Furthermore, we would expect $\xi_1$ to approach $\xi_0$ in the large core mass limit (nearly isothermal cores) -- the apparent lack of convergence is here attributed to numerical uncertainty.

\citet{Galli02} studied the structure and stability of MBESs by varying the external pressure and the strength of the ISRF (assuming outside shielding corresponding to $A_{\rm V} = 1$). They report different values for the density contrast than the ones presented here. In the particular case of $M = 5\,M_{\odot}$, they derive $\rho_{\rm c} / \rho_{\rm out} \sim 20$. This is an interesting result, as our models predict nearly isothermal values for the density contrast for high core masses. We attempted to reproduce their results by considering the density profile for the $M = 5\,M_{\odot}$ sphere as calculated here, but calculating the dust temperature assuming $A_{\rm V} = 1$. We calculated the gas temperature in the same way as in \citet{Galli02}, i.e. balancing heating and cooling functions using the parametrization of \citet{Goldsmith01}. We calculate $T_{\rm gas} \sim 12$\,K at the core center, decreasing to $\sim 9$\,K at about half radius. At the core edge, the temperature rises again to above $\sim 9.5$\,K. Using the gas temperature profile to solve the modified Lane-Emden equation (eq.\,\ref{eq2.3}) and solving the density contrast from eq.\,(\ref{rho_grad}) yields $\rho_{\rm c} / \rho_{\rm out} \sim 20$.

The high density contrast is probably due to the increased temperature at the core edge, where the increased pressure stabilizes the core. The parametrization of \citet{Goldsmith01} does not take into account the geometry at the edge of the core; in a more ``realistic'' scenario, increased photon escape probability at the core edge may allow gas temperatures to keep decreasing toward the core edge unless the cloud is so weakly shielded that the external UV field is able to heat its surface layers (through photoelectric effect). This scenario has recently been studied by \citet{Juvela11}. In the case of lower gas temperatures at the core edge, the predicted density contrast is probably closer to the isothermal value. However, proving this statement would certainly require a quantitative study.

Finally we note that while there is dependence on core mass, our models predict that the values of $\xi_1$ are equal to the isothermal critical value $\xi_0 \sim 6.45$ up to an accuracy of $\sim 5$\%. The density contrasts of the critical MBESs are also very similar to the density contrast of the critical BES, $\sim 14$.

\begin{figure}
\includegraphics[width=\columnwidth]{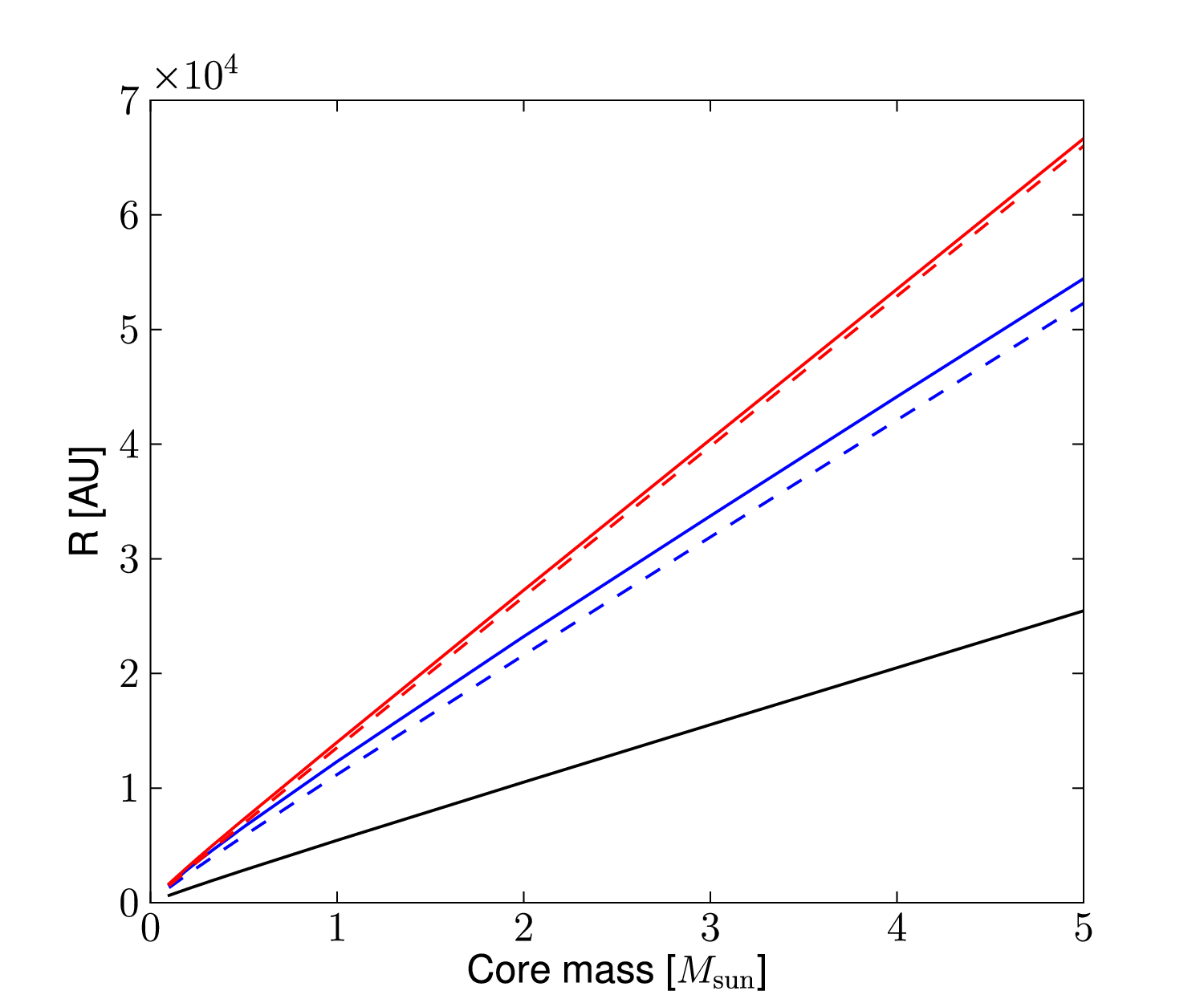}
\caption{The physical radii of Type 1 (blue solid line) and Type 2 (red solid line) critical cores as a function of core mass. The black curve represents homogeneous spheres in virial equilibrium (zero outside pressure) with temperature corresponding to the mean temperature of the 1 $M_{\odot}$ Type 1 MBES. Also plotted are the physical radii of critical isothermal cores corresponding to the mean temperature of each non--isothermal core (dashed lines; see text).}
\label{fig6}
\end{figure}

\subsection{Physical radii and boundary pressures} \label{sect3.2}

The physical radius, $R_{\rm out}$, of an MBES of given mass depends on the central temperature and the non-dimensional radius $\xi_{\rm out}$ according to
\begin{equation}
R_{\rm out} = \frac{GM}{c_{\rm s,c}^2} \,
\left[ {\xi_{\rm out} {\tau}_{\rm out}
\left(\frac{d\psi}{d\xi}\right)_{\rm out}} \right]^{-1} \; ,
\end{equation}
where $c_{\rm s,c} = kT_{\rm c}/m$ is the sound speed in the core centre.  For critical spheres, $\xi_{\rm out} = \xi_1$ is roughly constant (Fig.\,\ref{fig5}), and changes in the central temperatures are modest, so the radius is roughly proportional to the mass. 

We plot in Figure \ref{fig6} the physical radii of critical Type 1 (blue solid line) and Type 2 (red solid line) cores  as a function of core mass. The critical radii of the ``corresponding'' BESs with the temperature $T={\bar T}$, the average temperature of the MBES of the same mass, are plotted with dashed lines. The radii represent the {\sl minima} of the static equilibrium below which compression leads to the gravitational collapse. Also plotted is the physical radius of a virialized homogeneous, isothermal sphere with zero outer pressure, for which $R=GM/5c_{\rm s}^2$.  Also in this case we have used the assumption $T={\bar T}$.  One can see that the physical radii of MBESs are only slightly larger that those of the corresponding BESs. For the latter the critical radius can be approximated by $R\approx GM/2.4 c_{\rm s}^2$.

\begin{figure}
\includegraphics[width=\columnwidth]{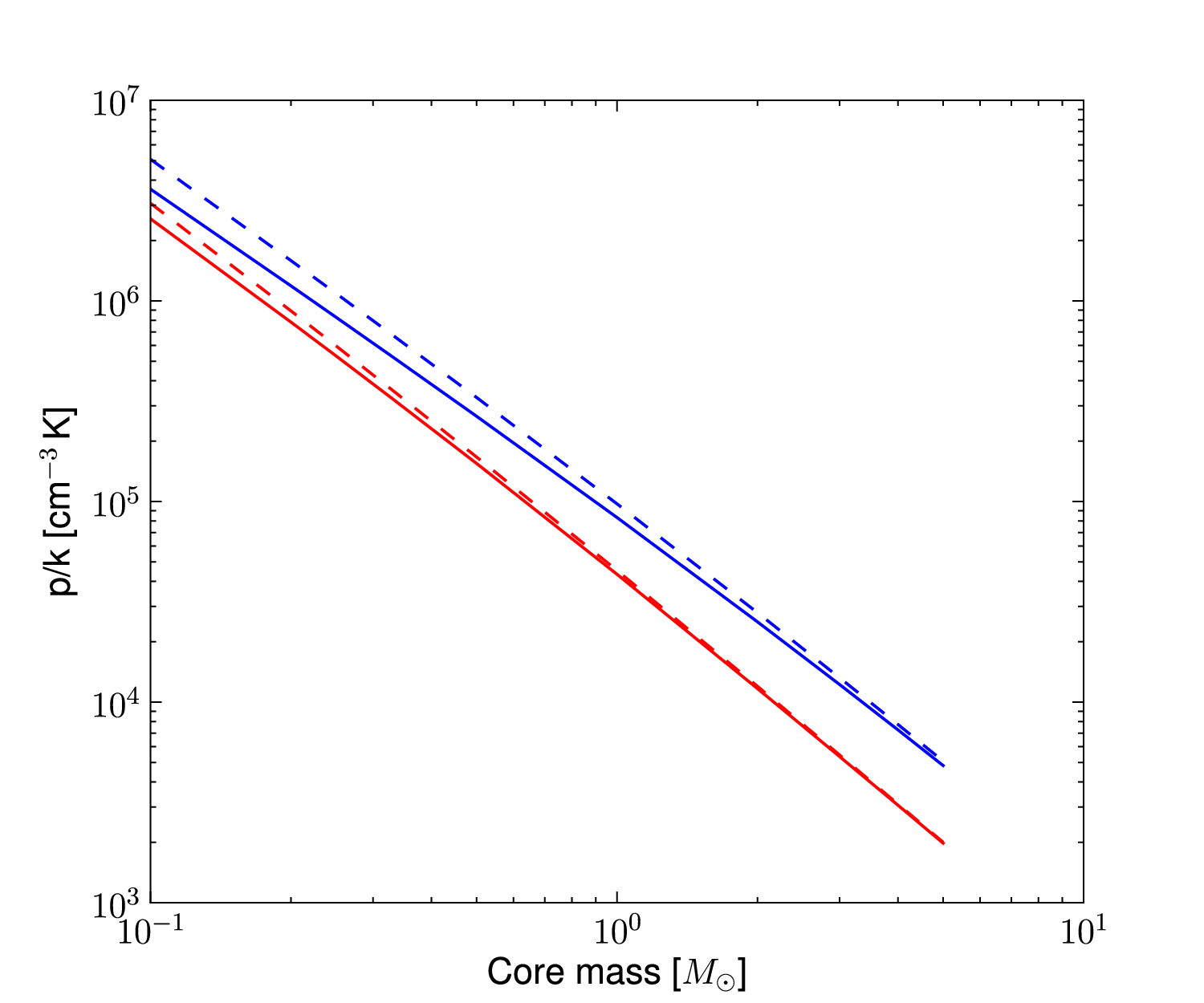}
\caption{The boundary pressures $p/k$ of critical Type 1 (blue line) and Type 2 (red line) MBESs as a function of core mass. Also plotted are the boundary pressures of BESs with temperatures corresponding to the mean temperatures of the MBESs (dashed lines).}
\label{fig7}
\end{figure}

To conclude this Section, we make a note about the boundary pressures of the critical MBESs. The boundary pressures $p/k$ of Type 1 and Type 2 cores as a function of core mass are plotted in Fig.\,\ref{fig7} (blue and red lines, respectively). The boundary pressures of the corresponding isothermal spheres are plotted with dashed lines. These diagrams represent the {\sl maximum} pressures above which an increase in the outer pressure leads to a gravitational collapse. For sub-critical cores, i.e. those with $\xi_{\rm out} < \xi_1$, the boundary pressure is lower and the physical radius is larger. The boundary pressures of critical Type 2 cores are smaller than those of Type 1 cores. Low-mass critical BESs have higher boundary pressures than the corresponding MBESs, but the difference decreases towards larger masses and is practically zero for $M = 5\,M_\odot$.

We have compared our calculations with those of \citet{Keto05}, who presented in their Fig.\,1 a p--V curve for an MBES of $M = 5\,M_{\odot}$. For the thermally supported sphere, \citet{Keto05} calculate for the critical sphere values of the gas density and boundary pressure of $\sim$\,10$^{4.3}$\,cm$^{-3}$ and $\sim$\,5.5$\times 10^3$\,cm$^{-3}$\,K, respectively. As also indicated by Figs.\,\ref{fig3} and \ref{fig7}, we predict\footnote{The central number density plotted in Fig.\,\ref{fig3} is the H$_2$ density, so that the {\sl total} hydrogen number density is twice that plotted in Fig.\,\ref{fig3}.} $\sim$\,10$^{4.1}$\,cm$^{-3}$ and $\sim$\,5$\times 10^3$\,cm$^{-3}$\,K for the central density and boundary pressure, respectively. Our results are thus comparable to those of \citet{Keto05}.

\section{Discussion} \label{sect4}

In this Section, we discuss the implications of the results presented in the previous Section. We make some general remarks about the nature of the stability and discuss the implications of the physical properties of the critical cores on the core chemistry. We conclude the Section with a discussion on very small, roughly Jupiter mass cores.

\subsection{The nature of the stability}\label{sect4.1}

\citet{Bonnor56} showed, by studying the second pressure derivative, that the minimum volume for a stable core corresponds to the configuration for which the pressure derivative with respect to volume $dp/dV$ at the core boundary becomes zero. The result applies here as well, although we consider non--isothermal cores. It follows (see also Sect.\,\ref{sect3.2}) that each physical radius shown in Fig.\,(\ref{fig6}) represents the smallest stable core configuration for a given core mass and model type. That is, for a given core mass, configurations with $R > R_1$ are stable, whereas {\sl all} configurations with $R < R_1$ are unstable.

At this point, however, it is prudent to consider what the stability actually means. The stability condition (eq.\,\ref{stability}) deals with small first--order fluctuations of the relevant parameters (central density, central temperature and non--dimensional radius). Thus in this context, the statement that a core is stable implies that the core is stable against small fluctuations of these parameters. The possibility that sufficiently large fluctuations could induce collapse even in a ``stable'' core cannot be ruled out based on the analysis presented here.

Also, our analysis does not give insight into what happens after a core begins to collapse. It may well be (as has been demonstrated in the literature and by our preliminary MHD collapse models) that ``unstable'' core configurations settle into a stable state relatively soon after collapse begins. This is due to the increase of the temperature inside the core: as the core collapses, the thermal pressure inside the core increases and may be able to support the slightly contracted core, depending on the core cooling mechanisms. Indeed, the stability condition presented here provides an initial condition for the collapse of non--isothermal cores, rather than trying to predict the ``absolute'' stability of the cores.

\subsection{Physical structures}\label{sect4.2}

Figures \ref{fig2} to \ref{fig4} demonstrate that a Type 1 MBES has a different physical structure than a Type 2 MBES of the same mass. In general, Type 1 cores are denser, warmer and smaller (Fig.\,\ref{fig6}) than Type 2 cores, i.e. Type 1 cores are more compact. Toward small core masses, the differences in radius and density diminish, but Type 1 cores always present a larger temperature gradient.

If one computes the mean temperature $\bar{T}$ of a given MBES, a ``corresponding'' BES of the same mass can be constructed. Figure \ref{fig3} shows that the central density of a BES constructed in this way is larger than that of the MBES -- this fact is also apparent in Fig.\,\ref{fig2} in the special case of $M = 0.25\,M_{\odot}$. The BES is, however, smaller. This result is general for small core masses. However, when using the gas temperature the situation changes to the opposite for larger masses \citep[see e.g.][]{Keto05}.

One can, of course, construct a BES with outer radius $R$ exactly equal to that of a given MBES. In this case, the temperature of the BES is clearly lower than the mean temperature of the MBES, but the density structures turn out to be nearly identical, with very small differences caused by the different temperature structures.

\subsection{Chemistry}\label{sect4.3}

We now briefly discuss the impact of the physical properties of the cores on their chemical evolution.

Let us first concentrate on the differences between the Type 1 and the Type 2 MBES. As discussed above, a Type 2 MBES is larger than a Type 1 MBES of the same mass, although the former is less dense and colder. The typical density difference in the center is however less than 15\% between the core types, so in this case, the major factor in the possible different chemical evolution of the cores is the temperature gradient.

Even small temperature changes can have significant consequences for the chemistry and the interpretation of observations. \citet{Aikawa05} studied the chemical evolution in collapsing clouds with initial conditions close to critical Bonnor-Ebert spheres.  Comparing models with a central density of $3\times 10^6$\,cm$^{-3}$ and temperatures in the range 10\,K--15\,K, the molecular depletion in the central parts of the core was found to be very sensitive to the temperature. In particular, the abundances of NH$_3$ and N$_2$H$^+$ could differ by an order of magnitude or more. However, because the central abundances are always low compared to the outer regions, it may be impossible to observationally confirm such a temperature dependence.

Let us then turn our attention to the differences between the MBES and the BES. As discussed in Sect.\,\ref{sect4.2}; the BES is always more compact than the corresponding MBES. The temperature dependence is more complex than when comparing two MBESs -- the BES is warmer in the center, but colder toward the edge. These differences may again produce notable differences in chemical evolution.

It is difficult to determine core temperature profiles from observations and, similarly, theoretical predictions contain significant sources of uncertainty. \citet{Goldsmith01} showed that molecular depletion can significantly increase the gas temperature predicted for cores. The effect can be several degrees even in dense clouds although, once the density is $\sim 10^5$\,cm$^{-3}$ or above, $T_{\rm gas}$ tends towards the dust temperature irrespective of the decreasing line cooling power. \citet{Juvela11} demonstrated further how, in addition to abundance variations, the radial temperature profiles can be modified by the velocity field (affecting the efficiency of line cooling) and the grain size distribution (affecting the gas-dust coupling). Therefore, each hydrostatic object is likely to exhibit slightly different density profiles.

The above implies that chemistry plays a role also in the dynamical evolution of a core. A study of the chemistry of the MBES including line cooling is planned -- this will allow a more realistic estimate of the gas temperature, also serving as a starting point for dynamical studies. In this context, one can also quantitatively study how the gas temperature affects the stability analysis presented here. Finally we note that the stability analysis presented in this paper indicates that an MBES can be just as stable as a BES of the same mass and (roughly) the same size, making it possible for advanced chemistries to develop in both cases in a typical core lifetime.

\subsection{Jupiter mass non--isothermal cores}

In the above, we present results of a stability analysis for core masses ranging from $0.1 \, M_{\odot}$ to $5.0 \, M_{\odot}$. It has been suggested that small dark clouds of roughly Jupiter mass could make up for part of the submm sources observed in e.g. SCUBA maps, or even account for part of the dark matter \citep{Lawrence01}. Although this scheme has later been found to be unlikely \citep[see e.g.][]{Drake03, Almaini05}, the existence of such objects is not entirely ruled out. For this reason, we thought it could be interesting to model these roughly Jupiter mass clouds (hereafter JMC) as either a BES or an MBES, to find out what kind of constraints would be imposed on the stability of these objects.

In our calculations thus far, we have assumed $A_{\rm V} = 10$ at the core edge, so that the objects are embedded in a bigger construct, e.g. a molecular cloud. However, if the JMCs are thought to be isolated, then the assumption of $A_{\rm V} = 10$ at the core edge is no longer justified. Indeed, if a JMC is isolated, the outermost parts are most likely ionized and heated to high temperatures by the ISRF, and molecular gas can be thought to be found at a minimum visual extinction of $A_{\rm V} \sim 2$. We have thus carried out the stability analysis for a core mass of $0.01\,M_{\odot}$ (roughly ten Jupiter masses), assuming outside $A_{\rm V} = 2$.

As a point of reference, let us first look at the BES. The critical radius of a 10\,$M_{\rm Jup}$ BES at 7\,K is 140\,AU, and the corresponding boundary pressure is $p/k = 4\times10^8$\,cm$^{-3}$\,K. As discussed in Sect.\,\ref{sect3}, these values represent the minimum physical radius and maximum boundary pressure for a core in stable equilibrium. The critical radius is directly proportional to the mass and the critical boundary pressure is proportional $M^{-2}$. So for example, a 1\,$M_{\rm Jup}$ BES at 7\,K would have a critical radius of 14\,AU and a boundary pressure of $4\times10^{10}$\,cm$^{-3}$\,K. According to \citet{Lawrence01}, the blank-field submm sources have angular radii of about $1 \arcsec$ or less, and if interpreted as cold dark clouds they should be nearby objects with a characteristic distance of 100\,pc. Hence a typical radius for this kind of cloud should be about 100\,AU. This radius roughly equals the critical radius of a 10 $M_{\rm Jup}$ BES. For a 1 $M_{\rm Jup}$ BES at 7\,K, the radius $R=100$\,AU corresponds to a sub-critical dimensionless radius of $\xi_{\rm out} \sim 1$. The boundary pressure for this configuration is still very high, $p/k \sim 2\times10^8$\,cm$^{-3}$\,K.

Let us then look at the MBES. For a critical Type 1 core, the physical radius is $\sim$\,200\,AU. The temperature gradient is rather steep (from 5\,K at the center to 10.5\,K at the edge), and the boundary pressure is $p/k \sim$\,$10^8$\,cm$^{-3}$\,K. The temperature gradient of the Type 2 core is shallower (from $\sim$\,4.8\,K at the centre to $\sim$\,8.5\,K at the edge), but $R$ and $p/k$ are nearly identical to those of the Type 1 core. This analysis suggests that the {\sl minimum} size of a stable 10\,$M_{\rm jup}$ MBES is $\sim$ 200\,AU, irrespective of the dust model used. However, the boundary pressure required to maintain the critical configuration in equilibrium is unlikely to be present in interstellar conditions, at least if the core is isolated. For clearly subcritical configurations ($\xi < \xi_1$, $R \sim 1000$\,AU), the boundary pressure is still $> 10^7$\,cm$^{-3}$\,K.

We conclude that due to the high boundary pressures required, roughly Jupiter mass critical BESs or MBESs are unlikely to be able to exist as isolated objects in interstellar space where the pressure is typically $\sim 10^4 -10^5$\,cm$^{-3}$\,K \citep[e.g.][]{McKee07}. Subcritical configurations can have radii of $\sim$\,1000\,AU, but the boundary pressure required for stable equilibrium is very high also for these.

\section{Conclusions} \label{sect5}

We studied the stability of non--isothermal Bonnor--Ebert spheres. The physical parameters of critical cores were derived for a range of core masses. Two different types of dust grains, corresponding to optical data from \citet{OH94} and \citet{LD01}, were considered in the modelling. The analysis was accordingly separated into two distinct core types (Type 1 and Type 2, respectively), each containing one type of dust grains.

As a general trend, the central density of a core increases as core mass decreases. The increase in density also increases the temperature gradient, because of more efficient attenuation of the external radiation field (in this paper, we study the dust temperature only). However, the absolute values of the temperature depend on the adopted dust model. Type 1 cores present larger temperature gradients than Type 2 cores, but Type 2 cores are colder. The difference also translates to core size: due to the larger temperature, the thermal pressure in the centres of Type 1 cores is larger and hence a Type 1 core can assume a smaller size than a Type 2 core of the same mass. Thus, Type 1 cores are more compact. Isothermal cores corresponding to non--isothermal cores (i.e. with same mass and same average temperature) are more centrally dense and slightly smaller than their non--isothermal counterparts. Considering gas temperature instead of dust temperature may affect core stability \citep[see e.g.][]{Keto05}. Our results should, however, hold well at least for cores with masses up to $\sim 0.5\,M_{\odot}$, for which the average density is $\sim 10^5$\,cm$^{-3}$.

It was found that the non--isothermal critical radius $\xi_1$ increases above the isothermal critical value $\xi_0 \sim 6.45$ toward small core masses. The effect is particularly pronounced for Type 1 cores, which present large temperature gradients. Nevertheless, the change in $\xi_1$ with core mass is rather small, and in the mass range studied here, $\xi_1$ is equal to $\xi_0$ to within 5\%. Furthermore, the ratio of central and outer densities for critical MBESs was found to present a similar increase toward small core masses, but the ratio is also in this case close to the isothermal value ($\sim 14$).

Although the differences in physical parameters between the two MBES types are not great, the different temperature gradients in particular may affect chemical evolution in these objects, which might be observable through line emission radiation from these objects. This may also the case when comparing isothermal and non--isothermal cores. Line radiation cooling can also affect the stability of the cores. A quantitative study of chemical evolution in the different types of cores could thus be justified.

We studied the physical parameters of MBESs of roughly ten Jupiter masses ($\sim 0.001-0.01\,M_{\odot}$) in order to validate or disqualify their possible existence as isolated objects. It was found that the boundary pressure required to maintain critical cores equilibrium is unlikely to be found in interstellar conditions. Even subcritical configurations require boundary pressures of $p/k \sim$\,$10^{7-8}$\,cm$^{-3}$\,K. We conclude that very low mass MBESs are unlikely to be able to exist in the interstellar medium as isolated objects -- this applies to very low mass BESs as well.

Finally, we note that the stability analysis presented in this paper considers small, first order variations of the relevant parameters (central density and temperature, core radius). Thus, stable cores as defined by the stability condition derived here are stable against linear perturbations, and sufficiently large perturbations could induce collapse even in these ``stable'' cores. Furthermore, what happens after collapse begins is not predicted by the equations. However, the near-critical gas spheres studied here
represent plausible, albeit idealistic models for cores at the very beginning of collapse.

\begin{acknowledgements}

O.S. acknowledges support from the Väisälä Foundation of the Finnish Academy of Science and Letters. The study has also been funded by the Academy of Finland through grants 132291 and 127015. The authors thank the referee Dr. Daniele Galli for helpful comments which improved the paper.

\end{acknowledgements}

\bibliographystyle{aa}
\bibliography{MBES.bib}

\appendix

\section{The stability condition} \label{appendixa}

In this Appendix we present a short derivation of the general form of the derivative of the boundary pressure with respect to core volume (eq.\,\ref{stability}). We also present an expression for the total number of particles in terms of $\beta$, $\lambda$ and $\xi_{\rm out}$. Throughout the calculations, we make use of equations (\ref{eq2.1}) to (\ref{eq2.3}) in the main text.

\subsection{The pressure derivative}

Calculating the variation of the pressure (eq.\,\ref{pressure}) with respect to the three non-dimensional variables $\beta$, $\lambda$ and $\xi_{\rm out}$ yields
\begin{eqnarray}
\delta p &=& {k \over m} \delta\left(\rho T\right) = 4\pi G \, \delta\left( \lambda  \beta  e^{-\psi} \right)  \nonumber \\
&=& 4\pi G  \beta \lambda e^{-\psi} \left[ \lambda^{-1} \delta\lambda + \beta^{-1} \delta\beta -  \left({d\psi \over d\xi}\right)_{\rm out} \delta\xi_{\rm out} \right] \, .
\end{eqnarray}
The variation of the cloud volume is
\begin{eqnarray}
\delta V &=& 4 \pi r^2 \delta r \nonumber \\
&=& 4\pi \beta^{3/2} \lambda^{-3/2} \xi_{\rm out}^2 \left[ {1\over2} \beta^{-1} \xi_{\rm out} \, \delta\beta -  {1\over2} \lambda^{-1} \xi_{\rm out} \delta\lambda + \delta\xi_{\rm out} \right] \, .
\end{eqnarray}
The above equations then yield the pressure derivative in general form:
\begin{equation}\label{eqb3}
{\delta p \over \delta V} = 2 G \beta^{-1/2} \lambda^{5/2} \xi_{\rm out}^{-3} e^{-\psi} { \left[ \beta^{-1} \delta\beta + \lambda^{-1} \delta\lambda - \left({d\psi \over d\xi}\right)_{\rm out} \delta\xi_{\rm out} \right] \over \left[ \beta^{-1} \delta\beta -  \lambda^{-1} \delta\lambda + 2 \xi_{\rm out}^{-1} \delta\xi_{\rm out} \right]  } \, .
\end{equation}
Identifying $2 G \beta^{-1/2} \lambda^{5/2} \xi_{\rm out}^{-3} e^{-\psi} = 2p/3V$ leads finally to eq.\,(\ref{stability}).

\subsection{Mass conservation}

We now present a non-isothermal version of the mass conservation equation discussed by \citet{Bonnor56}. Using eqs.\,(\ref{eq2.1}) and (\ref{eq2.2}), we can write the total number of particles as
\begin{equation}\label{eqbn}
N = {4\pi \over m} \int_0^{r} \rho y^2 dy = {4\pi \over m}  \, \beta^{3/2} \lambda^{-1/2} \int_0^{\xi_{\rm out}} { \xi^{2} \over \tau } e^{-\psi} d\xi \, ,
\end{equation}
The mass of the core is $M = Nm$. Since $N$ is calculated for a single core (that is, for a given value of $\xi_{\rm out}$), one can move $\beta$ and $\lambda$ out of the integral; for any single core, these are constant. In the integral, the variable $\xi$ represents the internal structure of the core and hence we make use of the solution to the Lane--Emden equation (eq.\,\ref{eq2.3}) that is {\sl unique to each core}. The above formula simplifies to
\begin{eqnarray}\label{eqbnumb}
N = {4\pi \over m} \beta^{3/2} \lambda^{-{1/2}} \, \xi^2 \tau_{\rm out} \left( {d\psi \over d\xi} \right) \, ,
\end{eqnarray}
where $\tau_{\rm out}$ represents the temperature contrast of the core, that is the ratio of outer and central temperatures. During the collapse process, we expect the total number of particles (i.e. the total mass of the core) to be conserved, i.e. that $\delta N = 0$. Calculation of this condition leads to the equation
\begin{equation}\label{eqb6}
\lambda^{-1} \delta \lambda = \beta^{-1} \delta\beta + {2 \over \tau^2} e^{-\psi} \biggl({d\psi \over d\xi}\biggr)^{-1} \delta \xi \, .
\end{equation}
It is readily verified that in the isothermal limit ($\delta\beta = 0$, $\tau = 1$), eq.\,(\ref{eqb6}) reduces to its isothermal analogue \citep[eq. 2.11 in][]{Bonnor56}.

\section{Thermodynamics of the MBES}\label{appendixb}

\citet{Bonnor56} discussed the form of Boyle's Law for a massive isothermal gas sphere. The radial change of the temperature causes a slight modification to Bonnor's equation of state, and we think there is a reason to make a brief excursion to the thermodynamics of an MBES. Besides writing down the equation of state, we also present the first law of thermodynamics in the case of an MBES which can be useful for dynamical calculations or when discussing thermal instability \citep[see][]{Keto05}.

It turns out that the internal energy of an MBES can be described by the simple ideal gas expression of internal energy when the temperature $T$ is replaced by the mass-averaged temperature $\bar{T}$ defined below. In general terms, an MBES can be considered as a thermodynamic system containing $N$ molecules in a volume $V$, at the temperature $\bar{T}$. The thermodynamic potentials and the equation of state can be derived much in the same way as for an ideal gas \citep[see e.g.][]{Landau69, Mandl88} from a partition function $Z$ which is the product of the perfect gas partition function, $Z_{\rm p}(N,V,{\bar T})$ (which may also contain rotational and vibrational parts), and the configurational partition $Z_{\Omega} = \exp({-\Omega/k{\bar T}})$, where $\Omega$ is the gravitational potential energy. For example, the Helmholtz free energy is then $F=-k{\bar T} \ln{Z}$, and the entropy can be obtained from $S = -\left(\frac{\partial F}{\partial {\bar T}}\right)_{N,V}$. Below, in the derivation of the internal energy, we use a more straighforward method. In any case, $\Omega$ has to be determined from the density distribution, and in this prescription it is a function of $N$ (mass), $V$ (outer radius), and an additional variable $\xi_{\rm out}$ which describes the distribution of mass within the volume $V$.

\subsection{Internal energy $U$}

Assuming that the medium consists of ideal gas with effective degrees of freedom\footnote{The very cold ($T \sim 10$\,K) ambient temperature in prestellar cores effectively removes the rotational degrees of freedom, so that even H$_2$ can be treated as a monatomic gas, at least to reasonable accuracy.} $f=3$, we can equate internal energy with total thermal energy \citep[see e.g.][]{Chandra57, Kippenhahn94} and write
\begin{equation} \label{eqb1.1}
U = 6\pi \int_0^R pr^2 dr \, .
\end{equation}
Using eqs.\,(\ref{eq2.1}) to (\ref{pressure}), we can write the integral in terms of $\beta$, $\lambda$, and $\xi$, and obtain
\begin{equation}\label{uglobal}
U = 24 \pi^2 G \, \beta^{5/2} \lambda^{-{1/2}} \int_0^{\xi_{\rm out}} \xi^{2} e^{-\psi} d\xi \, .
\end{equation}
On the other hand, the assumption of ideal gas means that the total internal energy of an isothermal core is
\begin{equation}\label{eq2.9}
U = {3\over2} Nk T \, .
\end{equation}
It can be shown that eq.\,(\ref{eq2.9}) is valid also for a non-isothermal sphere if $T$ is replaced with the average temperature ${\bar T}$ defined as
\begin{equation}\label{tw}
\bar{T} \equiv { \int_0^R T \rho \, dV \over \int_0^R \rho \, dV }  = {T_{\rm c} \over \xi_{\rm out}^{2} \tau_{\rm out} \left( {d\psi \over d\xi } \right)_{\rm out}  } \int_0^{\xi_{\rm out}} \xi^2 e^{-\psi} d\xi \, .
\end{equation}
When substituting the expressions of ${\bar T}$ and $N$ from eqs.\,(\ref{tw}) and (\ref{eqbn}) to eq.\,(\ref{eq2.9}) one obtains the formula for $U$ given in eq.\,(\ref{uglobal}).

\subsection{Gravitational potential energy $\Omega$ and the virial theorem} \label{sectb2}

\begin{figure}
\includegraphics[width=\columnwidth]{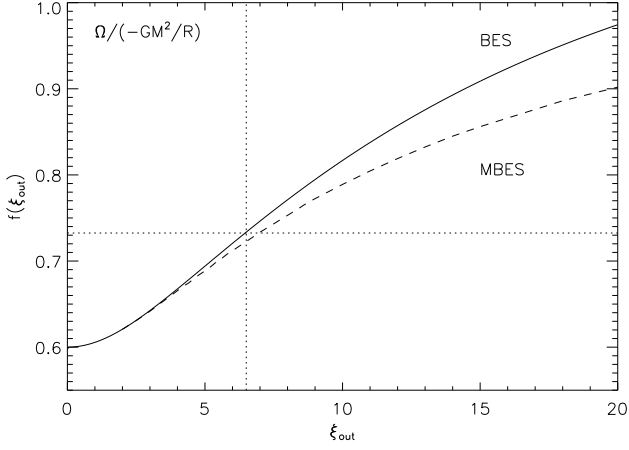}
\caption{The function $f(\xi_{\rm out})$ for a BES and an MBES of $M = 1\,M_{\odot}$.}
\label{figb1}
\end{figure}

The total potential energy of a gas sphere of radius $R$ can be written as
\begin{equation} \label{eqb2.1}
\Omega = -\int_0^{R} {GM \over r} dM = -4\pi \int_0^{R} {GM\rho \over r^2} r^3 dr \, ,
\end{equation}
where $\rho = \rho(r)$ and we have written the mass of a spherical shell as $dM = 4\pi r^2 \rho dr$. Using the hydrostatic equilibrium condition
\begin{equation}
{dp \over dr} = - {GM \rho \over r^2} \, ,
\end{equation}
we can rewrite eq.\,(\ref{eqb2.1}) into an easily integrable form:
\begin{equation} \label{eqb2.2}
\Omega = 4\pi \int_0^{R} {dp \over dr} r^3 dr = 4\pi p R^3 - 12\pi \int_0^{R} pr^2 dr \, .
\end{equation}
The latter term in eq.\,(\ref{eqb2.2}) can be identified as twice the thermal energy (see eq.\,\ref{eqb1.1}). Rearranging terms yields the {\sl virial theorem}
\begin{equation} \label{eqbvirial}
2 U + \Omega = 4\pi p R^3 = 3pV \, .
\end{equation}
Writing $\Omega$ in terms of $\beta$, $\lambda$ and $\xi_{\rm out}$ and using eqs.\,(\ref{eq2.2}) and (\ref{eq2.7}), it can be shown that
\begin{equation} \label{eqbov}
\Omega = - {GM^2 \over R} f(\xi_{\rm out}) = - \left( { 4\pi \over 3} \right)^{1/3} GM^2 V^{-1/3} f(\xi_{\rm out}) \, ,
\end{equation}
where $f$ is a function of $\xi_{\rm out}$ only. To illustrate, we plot in Fig.\,(\ref{figb1}) the ratio $f(\xi_{\rm out}) = \Omega / (-GM^2/R)$ for a BES and an MBES of $M = 1\,M_{\odot}$. Also marked is the isothermal critical value $\xi_0 \sim 6.45$. The Figure shows that the gravitational potential energies of the BES and the MBES differ from each other somewhat. For small values of $\xi_{\rm out}$ the cloud is
nearly homogenous and $f(\xi_{\rm out})$ approaches the value $0.6 = 3/5$.

\subsection{Equation of state}

Substituting the expression of the internal energy in terms of the average temperature ${\bar T}$ to the virial theorem (eq.\,\ref{eqbvirial}) one obtains the equation of state at the core boundary
\begin{equation}
pV = Nk{\bar T}  + \frac{\Omega}{3} \, .
\end{equation}
Using eq.\,(\ref{eqbov}), this can be written in the form suggested by \citet[][Bonnor's eq. 1.2]{Terletsky52}, criticized by Bonnor, if the numeric factor $\alpha$ is taken to be a function of $\xi_{\rm out}$.

\subsection{First law of thermodynamics}

Considering the internal energy only, the first law of thermodynamics states that
\begin{equation} \label{eqb11}
\delta U = T \delta S - p \delta V - { \partial \Omega \over \partial V} \delta V \, , 
\end{equation}
where the last term represents the work done by the gravitational potential energy. For a reversible process, the first term can replaced by heat $dQ$ supplied to the system. From eq.\,(\ref{eqbov}) it can be seen that
\begin{equation}
\frac{\partial \Omega}{\partial V} = - \frac{\Omega}{3V}  \, .
\end{equation}
Furthermore, using the equation of state the change of the internal energy
becomes
\begin{equation} \label{eqb13}
\delta U = {\bar T} \delta S - \frac{N k {\bar T}}{V} \delta V \, .
\end{equation}
The validity of this formula can be verified by using on the right hand side one of the Maxwell relations
\begin{equation}\label{eqbmaxwell}
\left( {\partial S \over \partial V} \right)_{\bar{T}, N} = \left( {\partial p \over \partial \bar{T}} \right)_{V, N} = {Nk \over V}  \, .
\end{equation}
On the other hand, from the definition of heat capacity:
\begin{equation}\label{eqbcv}
\left( {\partial S \over \partial \bar{T}} \right)_{V, N} = {1\over \bar{T}} C_V = {3 \over 2} {Nk \over \bar{T}} \, ,
\end{equation}
where we used the result that for a monatomic ideal gas $C_V~=~{3 \over 2} Nk$. Equations (\ref{eqbmaxwell}) and (\ref{eqbcv}) together yield
\begin{equation} \label{eqb16}
\bar{T} \delta S = {Nk\bar{T} \over V} \delta V + {3\over2} Nk \, \delta \bar{T} \, .
\end{equation}
Substituting this to the right hand side of eq.\,(\ref{eqb13}) one obtains the identity
\begin{equation}
\delta U = {3 \over 2} N k \, \delta {\bar T} \, ,
\end{equation}
in accordance with eq.\,(\ref{eq2.9}).

The total energy of the core equals to the sum of internal energy and gravitational potential energy, $E = U + \Omega$, from which follows that
\begin{equation} \label{eqbtotal}
\delta E = \delta U + \delta \Omega \, .
\end{equation}
According to eq. (\ref{eqbov}) we can write $\Omega = \Omega(V,\xi_{\rm out})$ when the mass is kept constant. Then
\begin{equation}\label{eqbomega}
\delta \Omega = {\partial \Omega \over \partial V} \delta V + {\partial \Omega \over \partial \xi_{\rm out}} \delta \xi_{\rm out} \, .
\end{equation}
Plugging eqs.\,(\ref{eqb11}) and (\ref{eqbomega}) to eq.\,(\ref{eqbtotal}) yields
\begin{equation}\label{eqbemid}
\delta E = \bar{T} \delta S - p \delta V + {\partial \Omega \over \partial \xi_{\rm out}} \delta \xi_{\rm out} \, .
\end{equation}
Using eq.\,(\ref{eqb16}) above the change of the total energy takes the form
\begin{equation}\label{eqbefinal}
\delta E = {3\over2} Nk \, \delta \bar{T}  - {\Omega \over 3V} \delta V + {\partial \Omega \over \partial \xi_{\rm out}} \delta \xi_{\rm out} \, .
\end{equation}
The above equation is another expression of the first law of thermodynamics.

\end{document}